# Delivery of Water and Volatiles to the Terrestrial Planets and the Moon[1]

M. Ya. Marov[a, *] and S. I. Ipatov[a, **]

[a]*Vernadsky Institute of Geochemistry and Analytical Chemistry, Russian Academy of Sciences, Moscow, 119991 Russia*
*e-mail: marovmail@yandex.ru
**e-mail: siipatov@hotmail.com

**Abstract** From modeling the evolution of disks of planetesimals under the influence of planets, it has been shown that the mass of water delivered to the Earth from beyond Jupiter's orbit could be comparable to the mass of terrestrial oceans. A considerable portion of the water could have been delivered to the Earth's embryo, when its mass was smaller than the current mass of the Earth. While the Earth's embryo mass was growing to half the current mass of the Earth, the mass of water delivered to the embryo could be near 30% of the total amount of water delivered to the Earth from the feeding zone of Jupiter and Saturn. Water of the terrestrial oceans could be a result of mixing the water from several sources with higher and lower D/H ratios. The mass of water delivered to Venus from beyond Jupiter's orbit was almost the same as that for the Earth, if normalized to unit mass of the planet. The analogous per-unit mass of water delivered to Mars was two−three times as much as that for the Earth. The mass of water delivered to the Moon from beyond Jupiter's orbit could be less than that for the Earth by a factor not more than 20.

## INTRODUCTION

The problem of the delivery of water and volatiles to the terrestrial planets is important for studying the origin and evolution of life in the Solar System and extrasolar systems (Marov et al., 2008; Marov, 2017). Liquid water is required for life to appear on planets. This problem is fundamental, since the Earth and terrestrial planets were formed in a high-temperature (~1000 K) zone of the protoplanetary disk, where water and volatiles are not retained, but accumulated beyond "the snow line" at a distance of $R > 3.5$ AU.

Endogenous and exogenous sources of water, as the main potential mechanisms forming terrestrial oceans, are considered in laboratory examinations of the materials and in computer simulations. In the laboratory, the terrestrial rocks are analyzed, while computer modeling is focused on the studies of a complex of the dynamical processes that occurred over the whole history of the Solar System. Both mechanisms have corresponding limitations, and we cannot exclude their mutual contribution to the solution of this problem.

The endogenous sources of water could include direct absorption of hydrogen from the nebula gas into the magmatic melts and a subsequent reaction of $H_2$ with FeO, which could increase the D/H ratio in the terrestrial oceans by a factor of 2−9 (Genda and Icoma, 2008), and accumulation of water by particles of the protoplanetary disk before gas began to dissipate in the inner part of the early Solar System (Drake and Campins, 2006; Muralidharan et al., 2008). The idea of a large amount of water in the mantle is confirmed by several studies; among them are the laboratory analysis of olivine in Archaean komatiite−basalt associations (ultramafic lavas in green belts of the Earth) produced in melting under extremal conditions at the boundary of the upper mantle of the Earth (Sobolev et al., 2016). The results of the study suggest that the mantle melts under a temperature of 1630 K and a fractional water content of ~0.5%, which corresponds to several terrestrial oceans, if extrapolated to the whole volume. The volume of water in minerals of the silicate Earth is estimated at 5−6 (to 50) volumes of the terrestrial oceans (Drake and Campins, 2006). A considerable amount of endogenous water may be contained in Mars and Venus. A limitation of the model is the formation of the Earth with temperatures higher than 500 K, which means that water and volatiles are not retained on the surface. Hallis et al. (2015) noted that the deep-mantle water has a small D/H ratio and could be acquired due to the water absorption by fractal particles during the accretion of the Earth.

---

[1] Reported at the Sixth International Bredikhin Conference (September 4–8, 2017, Zavolzhsk, Russia).

The water in oceans and its D/H ratio could have resulted from mixing the water from several sources with higher and lower D/H ratios.

The exogenous sources could originate from the migration of bodies from the outer part of the Main asteroid belt (O'Brien et al., 2014; Morbidelli et al., 2000, 2012; Petit et al., 2001; Raymond et al., 2004; Lunine et al., 2003, 2007) and migration of planetesimals from beyond Jupiter's orbit (Morbidelli et al., 2000; Levison et al., 2001; Marov and Ipatov, 2001; 2005; Ipatov and Mather, 2004, 2006, 2007; Ipatov, 2010). For the Grand Tack model, Rubie et al. (2015) considered the migration of planetesimals from a zone of 6−9.5 AU. In these scenarios, the authors estimated the probability of collisions of bodies with the Earth and other terrestrial planets and the mass of delivered water/volatiles. According to Drake and Campins (2006), a contribution of the bodies from beyond Jupiter's orbit to the water delivered to the Earth did not exceed ~50%.

Some authors believe that a substantial fraction of the water that came to the Earth from the outer asteroid belt. For example, Petit et al. (2001) thought that several embryos, which arrived from the outer asteroid belt at the end of Earth's formation, could deliver such amount of water to the Earth that is 10 times larger than the current amount of water on the Earth. O'Brien et al. (2014) supposed that water from the outer asteroid belt was mainly delivered by embryos as large as Mars. Drake and Campins (2006) noted that a key argument against the asteroid source of water, as the main source of water for the Earth, is that the isotopic composition of osmium (Os) of Earths primitive upper mantle matches that of anhydrous ordinary chondrites, not hydrous carbonaceous chondrites.

The model for the abundance of water and volatiles on the Earth (including the oceans and the atmosphere), which is based on migration of bodies from the outer Solar System, makes it possible to avoid the difficulties connected with the formation of terrestrial planets in a high-temperature zone of the protosolar disk. The model is limited by the difference between the D/H ratio for comets (excluding several comets, e.g., comet 103P/Hartley 2) and the standard value D/H = $1.5576 \times 10^{-4}$ for the terrestrial oceans (the Vienna Standard Mean Ocean Water (SMOW)). This limitation can be removed by the assumption that the main water source on the Earth was CI- and CM-chondrites rather than comets. Since, according to the investigations of lavas (Hallis et al., 2015), the deep-mantle water is likely to have a lower D/H ratio, it can be supposed that the present SMOW ratio is a result of some contribution of endogenous sources.

Pavlov et al. (1999) explained the deuterium-to-tritium paradox of the terrestrial oceans by the fact that the solar-wind-implanted hydrogen on dust particles provided the terrestrial oceans with a required fraction of water with a low D/H ratio. Delsemme (1999) believed that a large portion of ocean water was delivered by comets originated from Jupiter's zone, where vapor from the inner Solar System had condensed on interstellar ice grains before they accreted into large bodies. Drake and Campins (2006) suppose that the D/H and Ar/O ratios measured in cometary comas and tails do not adequately represent cometary interiors. Yang et al. (2013) showed that the D/H ratio for water is different for the bodies formed at different distances from the Sun. It was low for a hot inner disk, and then it increased with distance from the Sun and decreased again. Raymond and Izidoro (2017) think that C-type asteroids were formed at a distance of 5−20 AU from the Sun and passed to the current orbits when gas was still present in this zone. The characteristic time for the gas presence is estimated at 3−5 Myr (Zheng et al., 2017). As previously mentioned, some scientists believe that the supposition of the outer asteroid belt as the main source of water on the Earth explains the D/H ratio in the terrestrial oceans. However, if C-type asteroids came from the feeding zones of giants, as Raymond and Izidoro (2017) think, the water in the bodies, which arrived directly to the Earth from these zones, could also have the same D/H ratio as that in C-type asteroids and terrestrial oceans.

Our earlier studies of the delivery of water and volatiles to the terrestrial planets (e.g., Marov and Ipatov, 2001, 2005; Ipatov, 2010; Ipatov and Mather, 2004, 2006, 2007) were based on the results of numerical simulations of the migration of many thousands of small bodies and dust grains originating from such bodies; we considered the gravity effect of all of the planets for the case where the initial orbits of the bodies are close to the orbits of known comets and the masses and orbits of the planets take the current values.

The present analysis of the delivery of water and volatiles to the terrestrial planets from a zone beyond Jupiter's orbit is based on the results of our new calculations of the migration of planetesimals in the developing Solar System, and these simulations take into account the delivery of water to the growing terrestrial planets. They were also made for the embryos of the terrestrial planets rather than only for the planets themselves.

## INITIAL DATA AND ALGORITHMS TO MODEL THE MIGRATION OF SMALL BODIES

In our calculations, we modeled the migration of planetesimals under the gravity of planets. The current orbits and masses of the terrestrial planets, Jupiter, and Saturn were considered in the JS series of calculations. In the $JS_{01}$ series, the masses of terrestrial planets were 10 times smaller than their current values (in some cosmogonic models, it is assumed that Jupiter and Saturn were almost completely formed when the masses of terrestrial planets were far from their current values). In the JN and $JN_{01}$ series, Uranus and Neptune on their current orbits were additionally considered.

In four calculation series, JS, $JS_{01}$, JN, and $JN_{01}$, the semimajor axes $a$ of initial orbits of planetesimals were varied from $a_{min}$ = 4.5 to $a_{max}$ = 12 AU; and the number of planetesimals with a semimajor axis close to $a$ was proportional to $a^{1/2}$. The eccentricities and inclinations of initial orbits of planetesimals were 0.3 and 0.15 rad, respectively. As Ipatov (1993, 2000) noted, such eccentricities and inclinations could be reached due to the gravitational influence of planetesimals and planets. Two hundred and fifty planetesimals were usually considered in one calculation run, and the total number $N$ of planetesimals in a calculation series was 2000–2500.

We also considered the evolution of disks of planetesimals in the case where the giant planets (with their current masses) were located more closely to each other than at present (the maximum values of the semimajor axes of planetary orbits were varied from 15 to 20 AU) and $a_{max}$ did not exceed 23 AU. Specifically, in the $JN_{15}$ calculation series, the semimajor axes of initial orbits of the giants (with their current masses) were 5.45, 8.5, 12, and 15 AU, respectively, while the semimajor axes of the initial orbits of planetesimals were between 4.5 and 20 AU. Some runs (with 250 planetesimals) of these series of calculations (with closer orbits of giants, particularly, in the $JN_{15}$ calculation series) resulted in ejection of at least one of the giant planets (but not Jupiter) to a hyperbolic orbit in the course of evolution. Note that from observations of the microlensing events (Clanton and Gaudi, 2017), it is supposed that, on average, at least one free-floating exoplanet corresponds to one star.

To integrate the equations of motion, we used the symplectic method from the Swift package (Levison and Duncan, 1994). The gravitational influence of planets was taken into account. In different runs of calculations, the integration step was varied from 10 to 30 days and was constant in each of the runs. Earlier, we considered the evolution of orbits of more than 30000 bodies with initial orbits close to those of Jupiter-family comets (JFC), comet Halley, long-period comets and asteroids under the 3/1 and 5/2 resonances with Jupiter, and more than 20000 dust grains produced by these small bodies (Marov and Ipatov, 2005; Ipatov and Mather, 2004, 2006, 2007; Ipatov, 2010). In our previous calculations, we used the Bulirsch–Stoer algorithm (BULSTO) and the symplectic method of integration, which yielded almost the same results. In all of the considered series of calculations, when the initial orbits of bodies were close to those of several Jupiter-family comets, the probability $p_E$ of their collisions with the Earth during the dynamical evolution exceeded $4 \times 10^{-6}$, even if several bodies with the highest collision probability are excluded from the analysis. If the number of considered bodies is rather large, $p_E > 10^{-5}$. For the calculation series with the initial orbits that are close to the orbit of one of the comets, the values of $p_E$ for different comets could differ by a factor of almost 100. Among almost 30000 objects, whose initial orbits crossed Jupiter's orbit (the so-called Jupiter-crossing objects (JCOs)), several objects during the evolution acquired orbits lying completely within Jupiter's orbit and were moving along such orbits for millions or even hundreds of millions of years. The probability of collision of such an object with a terrestrial planet could be larger than the summed probability of thousands of other objects with almost the same initial orbits. The real objects that arrived from beyond Jupiter's orbit most likely break down to minicomets and dust grains during millions of years. However, the probability of falls of remnants of these objects on the planets is apparently not smaller than such a probability for the objects themselves. The probability that objects fall into the Sun did not exceed 0.02.

In the calculation runs for 250 planetesimals, the largest dynamical lifetime of planetesimals (till the moment when the distance of the last planetesimal to the Sun reaches 2000 AU or when it collides with the Sun) was varied from 0.9 to 3.9 Myr and from 5.9 to 47.2 Myr for the JS and JN series, respectively. The orbital elements of planetesimals obtained in our calculations for their dynamical lifetime were saved in the computer's memory and used to calculate the probability of planetesimal–planet collisions. For the JS, $JS_{01}$, JN, $JN_{01}$, and $JN_{15}$ calculation series, the probability of planetesimal–planet collisions during the dynamical lifetime of planetesimals are presented in Table 1. In the first line of Table 1, the data on 2250 planetesimals for the JS series are given. In the second line of this table, the probabilities are specified for 2550 planetesimals, including the previous 2250 pre-planetesimals. The main differences between these two lines were obtained for the probability of collisions with Mercury, since the additional calculation runs included the planetesimal with an orbit that crossed Mercury's orbit for a longer time than that in the other simulations.

On the basis of the obtained data arrays of the orbital elements for the dynamical lifetime of planetesimals, the probabilities of planetesimal–planet collisions were calculated. These probabilities were calculated not only for that mass of a planet on the Earth's orbit, which was used to model the evolution of disks of planetesimals, but also for a different mass of this planet. In simulations of the migration of planetesimals, the elements of their orbits were saved with a step of $d_t$ = 500 yr. For each set of the orbital elements of planetesimals and planets, the probabilities of collisions of planetesimals with a planet were calculated for an interval of 500 yr; and these probabilities were summed up over all sets of orbital elements.

**Table 1.** Probabilities $p_{pl}$ of collisions of planetesimals from the feeding zone of Jupiter and Saturn with different planets

|        | $N$  | Mercury | Venus | Earth | Mars | Jupiter | Saturn |
|--------|------|---------|-------|-------|------|---------|--------|
| JS     | 2250 | $1.58 \times 10^{-7}$ | $2.05 \times 10^{-6}$ | $2.07 \times 10^{-6}$ | $4.35 \times 10^{-7}$ | 0.048 | 0.0077 |
| JS     | 2550 | $7.39 \times 10^{-7}$ | $2.54 \times 10^{-6}$ | $2.62 \times 10^{-6}$ | $4.51 \times 10^{-7}$ | 0.047 | 0.0076 |
| JN     | 2000 | $0.92 \times 10^{-7}$ | $1.15 \times 10^{-6}$ | $1.92 \times 10^{-6}$ | $7.2 \times 10^{-7}$ | 0.041 | 0.006 |
| $JS_{01}$ | 2250 | $1.09 \times 10^{-7}$ | $2.35 \times 10^{-6}$ | $2.02 \times 10^{-6}$ | $4.49 \times 10^{-7}$ | 0.060 | 0.019 |
| $JN_{01}$ | 2000 | $1.32 \times 10^{-7}$ | $7.07 \times 10^{-7}$ | $1.11 \times 10^{-6}$ | $3.09 \times 10^{-7}$ | 0.041 | 0.0043 |
| $JN_{15}$ | 2550 | $6.11 \times 10^{-7}$ | $3.10 \times 10^{-6}$ | $4.52 \times 10^{-6}$ | $7.29 \times 10^{-7}$ | 0.186 | 0.031 |

**Table 2.** Relative probabilities of collisions of planetesimals with terrestrial planets. The quantities $p_{pl}/p_E$ and $p_{mE} = (p_{pl}/m_{pl})/(p_E/m_E)$, where $m_{pl}$ is the mass of a planet, $p_{pl}$ is the collision probability for a planetesimal and a planet, and $p_E$ and $m_E$ are the values of $p_{pl}$ and $m_{pl}$ for the Earth

|        | $N$  | Mercury | | Venus | | Earth | | Mars | |
|--------|------|---------|---|-------|---|-------|---|------|---|
|        |      | $p_{pl}/p_E$ | $p_{mE}$ | $p_{pl}/p_E$ | $p_{mE}$ | $p_{pl}/p_E$ | $p_{mE}$ | $p_{pl}/p_E$ | $p_{mE}$ |
| JS     | 2250 | 0.076 | 1.38 | 0.99 | 1.21 | 1 | 1 | 0.210 | 1.91 |
| JS     | 2550 | 0.282 | 5.10 | 0.97 | 1.19 | 1 | 1 | 0.172 | 1.56 |
| JN     | 2000 | 0.048 | 0.87 | 0.60 | 0.73 | 1 | 1 | 0.375 | 3.41 |
| $JS_{01}$ | 2250 | 0.0054 | 0.98 | 1.16 | 1.427 | 1 | 1 | 0.222 | 2.02 |
| $JN_{01}$ | 2000 | 0.0119 | 2.15 | 0.637 | 0.782 | 1 | 1 | 0.278 | 2.53 |
| $JN_{15}$ | 2550 | 0.0135 | 2.44 | 0.686 | 0.842 | 1 | 1 | 0.161 | 1.47 |

In calculations of the probability $p_{dts}$ of approaches of a planetesimal and a planet to the distance equal to the radius of the considered sphere $r_s$ (the sphere of action of a planet) for the time interval $d_t$, the following formulas were used in the 3D model (Ipatov, 2000): $p_{dts} = d_t/T_3$, where $T_3 = 2\pi^2 \cdot k_p \cdot T_s \cdot R \cdot k_v \cdot \Delta i \cdot R^2/(r_s^2 \cdot k_{fi})$ is the characteristic time before the encounter, $i$ is the angle (expressed in radians) between the orbital planes of the encountering celestial bodies, $R$ is the distance from the encounter location to the Sun, $k_{fi}$ is the sum of angles (expressed in radians) with vertices in the Sun, within which the distance between the projections of orbits onto the plane of the ecliptic is smaller than $r_s$, $T_s$ is the synodic period, $k_p = P_2/P_1$, $P_2 > P_1$, $P_i$ is the rotation period of the $i$th object (a planetesimal or a planet) about the Sun, $k_v = (2a/R - 1)^{1/2}$, and $a$ is the semimajor axis of the planetesimal's orbit (the coefficient $k_v$ was introduced by Ipatov and Mather (2004) to take into account the dependence of the encounter velocity versus the planetesimal's position on the eccentric orbit). The collision probability for the objects, which entered the sphere of action, was assumed to be $p_{dtc} = (r_\Sigma/r_s)^2(1 + (v_{par}/v_{rel})^2)$, where $v_{par} = (2Gm_\Sigma/r_\Sigma)^{1/2}$ is the parabolic velocity, $v_{rel}$ is the relative velocity of the objects coming to the distance $r_s$ of each other, $r_\Sigma$ is the sum of the radii of encountering objects with a total mass $m_\Sigma$, and $G$ is the gravitational constant. When the values $i$ are small, the formulas used in the algorithm were different. The algorithms (and their basis) to calculate $k_{fi}$ and the characteristic time between collisions of objects were described by S.I. Ipatov in Appendix 3 of Report O-1211 of the Keldysh Institute of Applied Mathematics of the Academy of Sciences of the Soviet Union for 1985 (p. 86−130). The probability of the planetesimal–planet collision $p_{dt}$ for the time $d_t$ is $p_{dts} \times p_{dtc}$. The values of $p_{dt}$ were summed up through the whole dynamical lifetime of a planetesimal.

Table 2 lists the values of $p_{pl}/p_E$ and $p_{mE} = (p_{pl}/m_{pl})/(p_E/m_E)$, where $m_{pl}$ is the mass of a planet, $p_{pl}$ is the probability of the planetesimal–planet collision, $p_E$ and $m_E$ are the quantities $p_{pl}$ and $m_{pl}$ for the Earth, respectively. The values of $p_{pl}/p_E$ characterize the ratios of the probability of collisions between planetesimals and terrestrial planets to the probability of collisions between planetesimals and the Earth. The values of the probability of collisions between planetesimals and a planet on the Earth's orbit $p_E$ and $p_{E01}$ (see Table 3) were calculated for the planetary mass equal to that of the Earth $m_E$ and $0.1 m_E$, respectively.

Table 4 presents the probabilities of collisions of planetesimals from the feeding zone of Jupiter and Saturn with a planet on the Earth's orbit $p_M$ and $p_{M01}$ for the planetary mass equal to that of the Moon $m_M$ and $0.1 m_M$, respectively (in the $JN_{15}$ series, $a_{max} = 20$ rather than 12 AU, as in the other calculation series). Since the orbits of the planetesimals crossing the Earth's orbit are strongly eccentric, the planetesimals moved in the sphere of action of the Earth to the lunar orbit (the semimajor axis of the lunar orbit is 2.4 times smaller than the radius of the sphere of action of the Earth and 3.8 times smaller than the Hill's radius of the Earth) along the trajectories slightly deviating from a rectilinear segment. Due to this, the calculated values of $p_M$ and $p_{M01}$ do not differ much from the real probabilities.

**Table 3.** Probabilities $p_E$ and $p_{E01}$ of collisions of planetesimals from the feeding zone of Jupiter and Saturn with the planets $m_E$ and $0.1 m_E$ in mass, respectively, on the Earth's orbit

|  | JS | JS | JS$_{01}$ | JN | JN$_{01}$ | JN$_{15}$ |
|---|---|---|---|---|---|---|
| $N$ | 2250 | 2550 | 2250 | 2000 | 2000 | 2550 |
| $p_E$ | $2.07 \times 10^{-6}$ | $2.62 \times 10^{-6}$ | $2.02 \times 10^{-6}$ | $1.92 \times 10^{-6}$ | $1.11 \times 10^{-6}$ | $4.52 \times 10^{-6}$ |
| $p_{E01}$ | $3.66 \times 10^{-7}$ | $4.70 \times 10^{-7}$ | $3.66 \times 10^{-7}$ | $3.32 \times 10^{-7}$ | $1.99 \times 10^{-7}$ | $8.24 \times 10^{-7}$ |
| $p_E/p_{E01}$ | 5.65 | 5.57 | 5.52 | 5.78 | 5.58 | 5.49 |
| $\log(p_E/p_{E01})$ | 0.752 | 0.746 | 0.742 | 0.762 | 0.746 | 0.740 |

**Table 4.** Probabilities $p_M$ and $p_{M01}$ of collisions of planetesimals from the feeding zone of Jupiter and Saturn with the planets $m_M$ and $0.1 m_M$ in mass, respectively, on the Earth's orbit

|  | JS | JS | JS$_{01}$ | JN | JN$_{01}$ | JN$_{15}$ |
|---|---|---|---|---|---|---|
| $N$ | 2250 | 2550 | 2250 | 2000 | 2000 | 2550 |
| $p_M$ | $1.24 \times 10^{-7}$ | $1.59 \times 10^{-7}$ | $1.21 \times 10^{-7}$ | $1.16 \times 10^{-7}$ | $6.74 \times 10^{-8}$ | $2.71 \times 10^{-7}$ |
| $p_{M01}$ | $2.64 \times 10^{-8}$ | $3.39 \times 10^{-8}$ | $2.58 \times 10^{-8}$ | $2.49 \times 10^{-8}$ | $1.44 \times 10^{-8}$ | $5.75 \times 10^{-8}$ |
| $p_E/p_M$ | 16.70 | 16.50 | 16.72 | 16.58 | 16.47 | 16.68 |
| $p_M/p_{M01}$ | 4.70 | 4.68 | 4.69 | 4.66 | 4.68 | 4.71 |
| $\lg(p_M/p_{M01})$ | 0.672 | 0.670 | 0.670 | 0.670 | 0.670 | 0.673 |

## CALCULATION RESULTS FOR THE PLANETESIMALS' MIGRATION

In our previous calculations (Ipatov and Mather, 2004, 2006, 2007), we considered the migration of bodies, whose initial orbits were close to those of comets crossing Jupiter's orbit. For the current orbits and masses of the planets, those calculations yielded the values of $p_E$ higher than $4 \times 10^{-6}$. In new JS and JN series of calculations, the probability $p_E$ of a planetesimal–Earth collision is approximately $2 \times 10^{-6}$. This value was obtained in the consideration of thousands of planetesimals. In different runs with 250 initial planetesimals, the values of $p_E$ may differ by an order of magnitude for the same series of calculations. For planetesimals that initially were in the inner part of the disk, $p_E > 2 \times 10^{-6}$. In the calculations made by Morbidelli et al. (2000) for planetesimals that initially were on circular orbits with zero inclinations, the values of $p_E$ were around $(1–3) \times 10^{-6}$ for the semimajor axes of orbits ranging from 5 to 8 AU. In our JS, JN, and JS$_{01}$ calculation series, the collision probability $p_{E01}$ for a planetesimal and the Earth's embryo $0.1 m_E$ in mass was estimated at approximately $4 \times 10^{-7}$.

In the Grand Tack model, the region between 3 and 6 AU was considered to be cleared of planetesimals due to the migration of Jupiter toward the Sun and back (Rubie et al., 2015). In the paper by Raymond and Izidoro (2017), it is affirmed that, after the formation of Jupiter, most planetesimals left a zone between 4 and 7 AU during approximately 1 Myr, when gas was still present there. If a lower boundary of the disk is assumed at 6 AU instead of 4.5 AU used in our calculations, the obtained value of $p_E$ can be some-what smaller than that in our simulations.

In the JS, JS$_{01}$, JN, and JN$_{01}$ calculation series, a portion of planetesimals reaching the Earth's orbit was 12–14%. If $p_E$ is calculated only for such planetesimals, $p_E$ will be almost an order of magnitude higher than $2 \times 10^{-6}$. In the calculation series with initially close mutual positions of the giant planets, the values of $p_E$ and $p_{E01}$ were mostly not smaller than those for the JS, JS$_{01}$, JN, and JN$_{01}$ series. Specifically, $p_E \approx 4.5 \times 10^{-6}$ in the JN$_{15}$ series, where 2/3 of the initial planetesimals had the semimajor axes larger than 12 AU. The larger values of $p_E$ are caused by the more intense migration of planetesimals toward the Earth's orbit. The stronger migration of planetesimals inward the Solar System could be obtained from the consideration of the mutual gravitation influence of planetesimals, which was ignored in our simulations. Because of this, in the real Solar System, the probability of collisions of planetesimals, which came from the feeding zone of the giants, with the Earth could be of the order of $4 \times 10^{-6}$.

## DELIVERY OF WATER TO TERRESTRIAL PLANETS

From the assumption that $p_E = 2 \times 10^{-6}$ and the total mass of planetesimals in the feeding zone of Jupiter and Saturn was roughly 100 Earth masses (Ipatov, 1993, 2000), while the fraction of water $k_W$ in planetesimals was 0.5, we find that the total mass of water delivered from this zone to the Earth could be approximately $10^{-4} m_E$ (about $6 \times 10^{20}$ kg), i.e., roughly half the water mass in the terrestrial oceans (the latter is $1.4 \times 10^{21}$ kg). Moreover, almost the same amount of water could be delivered to the Earth from a zone that is beyond 12 AU from the Sun. Planetesimals could mostly migrate from this zone later than from the feeding zone of Jupiter and Saturn; and a substantial portion of water, which came from beyond Saturn's orbit, could fall on the Earth's embryo when its mass was not small. Accounting for the mutual gravitational influence of planetesimals leads to the increase in the orbital eccentricities of many planetesimals, the portion of planetesimals reaching the Earth's orbit, and the probability of collisions of planetesimals with the Earth and other terrestrial planets. The total mass of water delivered to the Earth from beyond Jupiter's orbit could be comparable to that in the terrestrial oceans.

In the above estimates of the amount of water delivered to the Earth, the total mass of bodies beyond Jupiter's orbit was approximately determined as 200 Earth masses. The disk of planetesimals with a mass up to $200 m_E$ was also considered by Hahn and Malhotra (1999). Morbidelli et al. (2012) assumed this mass to be equal to 35–50 Earth masses, while the contribution of such bodies to the terrestrial oceans was estimated at 10%. In support of the hypothesis of a probable large total mass of planetesimals beyond Jupiter's orbit, we note that, in our calculations, the fraction of planetesimals that experienced collisions with Saturn was essentially less than 1%, while it was even smaller for Uranus and Neptune. Because of this, for each body that encountered these three planets, there were dozens of bodies ejected to hyperbolic orbits, while the total mass of only Uranus and Neptune exceeds $30 m_E$.

If the estimates of the water fraction in planetesimals are lower, the estimates of water delivered to the Earth from beyond Jupiter's orbit are smaller. Morbidelli et al. (2012) and Marty et al. (2016) noted that the fraction of water in planetesimals did not exceed 50%. Rubie et al. (2015) believed that the fraction of water ice in the bodies formed at a distance larger than

6 AU was 20%. According to Greenberg (1998), in a cometary nucleus, the water fraction is about 30%. Davidsson et al. (2016) came to conclusion that the fraction of ice in comet 67P is within the limits from 14 to 33%. Some authors believe that primary planetesimals could contain more ice than comets in our day. Fulle et al. (2017) suppose that, though the volume fraction of water in comet 67P and trans-Neptunian objects is approximately 20%, the bodies born close to the snow line contained more water than trans-Neptunian objects.

If the loss of water in collisions of planetesimals with the Earth is taken into account, the estimate of a water portion delivered to the Earth from beyond Jupiter's orbit decreases. Canup and Pierazzo (2006) found that, if a planetesimal collides with the Earth with a velocity which is higher than the parabolic velocity by more than 1.4 times and the collision angle is larger than 30°, more than 50% of impactor's water is lost.

In the runs presented in Table 2, the ratio $p_{pl}/m_{pl}$ of the probability of planetesimal–planet collisions to the planetary mass calculated for Mars, Venus, and Mercury is higher than that for the Earth by 1.5–3.4, 0.7–1.4, and 0.9–5.1 times, respectively. These estimates suggest that the mass of planetesimals or water delivered to Venus from beyond Jupiter's orbit was approximately the same as that for the Earth, if taken per unit mass of the planet; at the same time, the analogous mass of planetesimals or water delivered to Mars was 2−3 times larger than that for the Earth, if taken per unit mass of the planet. In absolute value, the water mass delivered to Mars from beyond Jupiter's orbit was 3−5 times smaller than that delivered to the Earth. For Mercury, the ratio $p_{mE} = (p_{pl}/m_{pl})/(p_E/m_E)$ was not smaller than that for the Earth. These values of the probability of planetesimal–planet collisions are consistent with our earlier estimates made for the objects, the initial orbits of which crossed Jupiter's orbit (Ipatov and Mather, 2004, 2006, 2007). The estimates are indicative of the presence of ancient oceans on Mars and Venus, which could partially survive deep under the surface (as on Mars (Usui, 2017; Wade, 2017)) or were lost in the course of evolution (as on Venus (Kasting, 1988; Marov, Grinspoon, 1998, chapter 9; Chassefière et al., 2012; Marov, 2017, p. 145–147)).

## FALL OF PLANETESIMALS ONTO A GROWING EMBRYO OF THE EARTH

From the arrays of the orbital elements of planetesimals and planets calculated for different times, the probabilities of collisions of planetesimals or comets with a planet on the Earth's orbit were determined for the mass of the planet equal to $m_E$ and $0.1 m_E$; and the ratio of probabilities $p_E/p_{E01}$ was found in the range from $5.5 \approx 10^{0.74}$ to $5.8 \approx 10^{0.76}$. Thus, the ratio of the mass of planetesimals falling onto the planet to the mass of the planet is approximately two times higher for the planetary mass $0.1m_E$ than for $m_E$. Here, we consider the planetesimals that came from the feeding zone of giant planets (let us call them $j$-planetesimals). For the planetesimals that came from the feeding zone of terrestrial planets, the index of power is larger than 1; i.e., the larger planets grew faster.

If we take into account that the effective radius of the planet of radius $r$ is approximately $r_{eff} \approx r(1 + (v_{par}/v_{rel})^2)^{1/2}$ and the parabolic velocity on the planetary surface $v_{par}$ is proportional to $r^{-1/2}$, we can find the relative velocity of a planetesimal entering the sphere of action of the planet $v_{rel} \approx 11.2(1 - 10^{-5/3}(p_E/p_{E01}))^{1/2}/(10^{-2/3}(p_E/p_{E01}) - 1)^{1/2}$ km/s from the ratio $p_E/p_{E01} = (r_{effE}/r_{effE01})^2$ (where $r_{effE}$ and $r_{effE01}$ are the effective radii of the Earth and its embryo of mass $0.1m_E$, respectively). In particular, $v_{rel}$ is 21.0, 23.1, and 24.4 km/s, if $p_E/p_{E01}$ is 5.8, 5.6, and 5.5 respectively. For comparison, according to several models by Nesvorný et al. (2017), asteroids, which initially were on the orbits with semimajor axes ranging from 1.6 to 3.3 AU, have mean velocities of collisions with the Earth varying from 21 to 23.5 km/s.

If the effective radius of the body is close to its radius, the effective cross-section of the body of mass $m$ is roughly proportional to $m^{2/3}$. For such a model, the power index is $2/3 \approx 0.667$, which is slightly smaller than the power indices obtained in the analysis of collisions of $j$-planetesimals with the Earth. The ratio of the effective cross-section (proportional to $m^{2/3}$) to the mass $m$ is proportional to $m^{-1/3}$; i.e., in this case, the relative growth of the planetary mass is more rapid for less massive planets. For weakly eccentric orbits, on the contrary, the larger bodies grow quicker.

If the relative mass increase of an embryo at the expense of $j$-planetesimals is proportional to $m^{0.74}$, the ratio of the embryo's mass increase from 0 to $km_E$ to that from 0 to $m_E$ is $k^{1.74}$. And, $0.5^{1.74} \approx 0.3$ and $0.8^{1.74} \approx 0.68$. The fraction of $j$-planetesimals, which fell onto the embryo during its increase in mass to $km_E$, may be smaller than $k^{1.74}$ if the ratio of the inflow of $j$-plane-tesimals to that of "local" planetesimals at terminal stages of the planet formation was larger than such a ratio at early stages of the embryo growth. With the above estimates of the material migration from beyond Jupiter's orbit to the Earth, we may find that, when the Earth's embryo was growing to $0.5m_E$, the mass of water delivered to the embryo could be around 30% of all the water delivered from the feeding zones of Jupiter and Saturn. The above estimates show that a substantial mass of water could be delivered to the Earth's embryo when its mass was smaller than the present mass of the Earth. In the Grand Tack model, most bodies which originated from the zone beyond 6–7 AU fell onto the Earth after the latter possessed 60–80% of its final mass (Rubie et al., 2015).

## DELIVERY OF WATER AND VOLATILES TO THE MOON

In the calculation series presented in Table 4, the probability of collisions of planetesimals with the Moon varied from $7 \times 10^{-8}$ to $2.7 \times 10^{-7}$; and, in three calculation series, they were approximately $1.2 \times 10^{-7}$. In the considered calculation series, the ratio of the probability of collisions of planetesimals with the Earth to that with the Moon $p_E/p_M$ was within the limits from 16.5 to 16.7. In the JS and JN calculation runs with $N = 250$ planetesimals, the ratio $p_E/p_M$ varied from 16.53 to 16.9 and from 16.08 to 16.74, respectively. In the calculation runs, where the initial positions of the giant planets were assumed to be closer (in particular, $JN_{15}$), the ratio $p_E/p_M$ varied from 16.0 to 17.0. By comparison, the squared ratio of the radii of the Earth and the Moon is 13.48. In different runs with $N = 250$ in the same calculation series, the probabilities of collisions with the Moon (or any planet) could differ by more than 10 times. The mass of planetesimals and water delivered to the Moon from beyond Jupiter's orbit could be smaller than that for the Earth by the factor not more than 20.

For migrating objects, the initial orbits of which crossed Jupiter's orbit and were close to those of Jupiter-family comets, the ratios $p_E/p_M$ were also calculated in different runs (with 250 objects) (Ipatov and Mather, 2004, 2006, 2007). These values varied from 15.2 to 17.6. For asteroids from the 3:1 resonance with Jupiter and comets with eccentricity $e = 0.975$, $p_E/p_M$ reached 18.6 and 15.2, respectively. In these simulations, the scattering in the $p_E/p_M$ values was from 5.1 to 6.0. The ratio of the probability of collisions with the Moon with its present density to that with its density equal to that of the Earth was close to 1.39 in all calculation runs.

For comets from the trans-Neptunian belt (with initial distances in a range of 20–30 AU), Nesvorný et al. (2017) found that the probabilities of collisions with Venus, the Earth, Mars, and the Moon are $3.7 \times 10^{-7}$, $5.0 \times 10^{-7}$, $9.1 \times 10^{-8}$, and $2.6 \times 10^{-8}$, respectively. For asteroids with initially large semimajor axes, from 1.6 to 3.3 AU, these probabilities were $(1.2–1.5) \times 10^{-2}$, $(1.0–1.1) \times 10^{-2}$, $(3.6–3.9) \times 10^{-3}$, $(4.4–5.3) \times 10^{-4}$, respectively. From the data reported in the abovementioned paper, the ratio $p_E/p_M$ can be estimated as 19 and 21–23 for comets and asteroids, respectively.

## CONCLUSIONS

The evolution of disks of planetesimals under the influence of planets was simulated. The results of calculations showed that, in the course of the formation of the Solar System, the mass of water delivered to the Earth from beyond Jupiter's orbit could be comparable to the mass of terrestrial oceans. While the Earth's embryo was growing to half the present mass of the Earth, the mass of water delivered to the embryo could be approximately 30% of all the water delivered to the Earth from the feeding zone of Jupiter and Saturn. The water in terrestrial oceans and its D/H ratio could result from the mixing of water from several sources with high and low D/H ratios. The mass of water delivered to Venus from beyond Jupiter's orbit was approximately the same as that for the Earth, if taken per unit mass of the planet; at the same time, the analogous per-unit mass of water delivered to Mars was 2−3 times larger than that for the Earth. In absolute value, the water mass delivered to Mars from beyond Jupiter's orbit was 3−5 times smaller than that delivered to the Earth. The mass of water delivered to Mercury was not smaller than that for the Earth, if taken per unit mass of the planet. The mass of water delivered to the Moon from beyond Jupiter's orbit could be smaller than that for the Earth by the factor not more than 20.

## ACKNOWLEDGMENTS

The studies of the material migration to the Moon were supported by the Russian Scientific Foundation (project no. 17-17-01279), and the studies of the delivery of water and volatiles to the terrestrial planets were supported by the Program 17 of the Presidium of the Russian Academy of Sciences (government contract no. 00137-2018-0030 of the Vernadsky Institute of Geochemistry and Analytical Chemistry of the Russian Academy of Sciences).

## REFERENCES

Canup, R.M. and Pierazzo, E., Retention of water during planet-scale collisions, *Proc. 37th Lunar and Planetary Sci. Conf.,* League City, TX, 2006, abs. #2146.

Chassefière, E., Wieler, R., Marty, B., and Leblanc, F., The evolution of Venus: present state of knowledge and future exploration, *Planet. Space Sci.,* 2012, vol. 63–64, pp. 15–23.

Clanton, C. and Gaudi, B.S., Constraining the frequency of free-floating planets from a synthesis of microlensing, radial velocity, and direct imaging survey results, *Astrophys. J.,* 2017, vol. 834, art. ID A46, 13 p.

Davidsson, B.J.R., Sierks, H., Güttler, C., Marzari, F., Pajola, M., Rickman, H., A'Hearn, M.F., Auger, A.-T., El-Maarry, M.R., Fornasier, S., Gutierrez, P.J., Keller, H.U., Massironi, M., Snodgrass, C., Vincent, J.-B., and 33 co-authors, The primordial nucleus of comet 67P/Churyumov-Gerasimenko, *Astron. Astrophys.,* 2016, vol. 592, art. ID A63, 30 p.

Delsemme, A.H., The deuterium enrichment observed in recent comets is consistent with the cometary origin of seawater, *Planet. Space Sci.,* 1999, vol. 47, pp. 125–131.

Drake, M. and Campins, H., Origin of water on the terrestrial planets, *Proc. 229th IAU Symp. "Asteroids, Comets, and Meteors",* 2006, pp. 381–394.

Fulle, M., Della Corte, V., Rotundi, A., Green, S.F., Accolla, M., Colangeli, L., Ferrari, M., Ivanovski, S., Sordini, R., and Zakharov, V., The dust-to-ice ratio in comets and Kuiper belt objects, *Mon. Notic. Roy. Astron. Soc.,* 2017, vol. 469, pp. S45–S49.

Genda, H. and Icoma, M., Origin of the ocean on the Earth: early evolution of water D/H in a hydrogen-rich atmosphere, *Icarus,* 2008, vol. 194, no. 1, pp. 42–52.

Greenberg, J.M., Making a comet nucleus, *Astron. Astrophys.,* 1998, vol. 330, pp. 375–380.

Hahn, J.M. and Malhotra, M., Orbital evolution of planets embedded in a planetesimal disk, *Astron. J.,* 1999, vol. 117, pp. 3041–3053.

Hallis, L.J., Huss, G.R., Nagashima, K., Taylor, G.J., Halldórsson, S.A., Hilton, D.R., Mottl, M.J., and Meech, K.J., Evidence for primordial water in Earth's deep mantle, *Science,* 2015, vol. 350, pp. 795–797.

Ipatov, S.I., Migration of bodies in the accretion of planets, *Sol. Syst. Res.,* 1993, vol. 27, no. 1, pp. 65–79.

Ipatov, S.I., *Migratsiya nebesnykh tel v Solnechnoi sisteme (Celestial Bodies' Migration in Solar System),* Moscow: URSS, 2000. 320 p. http://www.rfbr.ru/rffi/ru/books/ o_29239, http://booksee.org/book/1472075.

Ipatov, S.I., Collision probabilities of migrating small bod-ies and dust particles with planets, *Proc. Int. Astron. Union, Symp. S263. "Icy Bodies in the Solar System" (Rio de Janeiro, Brazil, Aug. 3–7, 2009),* Fernandez, J.A., Lazzaro, D., Pri-alnik, D., and Schulz, R., Eds., Cambridge Univ. Press, 2010, pp. 41–44. http://arxiv.org/abs/0910.3017.

Ipatov, S.I. and Mather, J.C., Comet and asteroid hazard to the terrestrial planets, *Adv. Space Res.,* 2004, vol. 33, no. 9, pp. 1524–1533. http://arXiv.org/format/astro-ph/ 0212177.

Ipatov, S.I. and Mather, J.C., Migration of small bodies and dust to near-Earth space, *Adv. Space Res.,* 2006, vol. 37, no. 1, pp. 126–137. http://arXiv.org/format/ astro-ph/0411004.

Ipatov, S.I. and Mather, J.C., Migration of comets to the terrestrial planets, in *Proc. IAU Symp. no. 236 "Near-Earth Objects, Our Celestial Neighbors: Opportunity and Risk" (Aug. 14–18, 2006, Prague, Czech Republic),* Milani, A., Valsecchi, G.B., and Vokrouhlický, D., Eds., Cambridge: Cambridge Univ. Press, 2007, pp. 55–64. http://arXiv.org/format/astro-ph/0609721.

Kasting, J.F., Runaway and moist greenhouse atmospheres and the evolution of Earth and Venus, *Icarus,* 1988, vol. 74, pp. 472–494.

Levison, H.F. and Duncan, M.J., The long-term dynamical behavior of short-period comets, *Icarus,* 1994, vol. 108, no. 1, pp. 18–36.

Levison, H.F., Dones, L., Chapman, C.R., Stern, S.A., Duncan, M.J., and Zahnle, K., Could the lunar "late heavy bombardment" have been triggered by the formation of Uranus and Neptune?, *Icarus,* 2001, vol. 151, pp. 286–306.

Lunine, J.I., Chambers, J., Morbidelli, A., and Leshin, L.A., The origin of water on Mars, *Icarus,* 2003, vol. 165, no. 1, pp. 1–8.

400  MAROV, IPATOV

Lunine, J., Graps, A., O'Brien, D.P., Morbidelli, A., Leshin, L., and Coradini, A., Asteroidal sources of Earth's water based on dynamical simulations, *Proc. 38th Lunar and Planetary Sci. Conf.,* League City, TX, 2007, abs. #1616.

Marov, M.Ya. and Grinspoon, D.H. *The planet Venus*, New Haven: Yale University Press, 1998, 442 p.

Marov, M.Ya. and Ipatov, S.I., Volatile inventory and early evolution of planetary atmospheres, in *Collisional Processes in the Solar System,* Marov, M.Ya. and Rickman, H., Eds., Dordrecht: Kluwer Acad. Publ., 2001, pp. 223– 247.

Marov, M.Ya. and Ipatov, S.I., Migration of dust particles and volatiles delivery to the terrestrial planets, *Solar Syst. Res.,* 2005, vol. 39, no. 5, pp. 374–380.

Marov, M.Ya., Kolesnichenko, A.V., Makalkin, A.B., Dorofeeva, V.A., Ziglina, I.N., and Chernov, A.V., From proto-Sun cloud to the planetary system: gas-dust disk evolution model, in *"Problemy zarozhdeniya i evolyutsii biosfery"* (Problems on Biosphere Origin and Evolution), Galimov, E.M., Ed., Moscow: LIBRO-KOM, 2008, pp. 223–273, in Russian.

Marov, M.Ya., *Kosmos. Ot solnechnoi sistemy vglub' Vselennoi (Space. From the Solar System deep into the Universe)*, Moscow: Fizmatlit, 2017, 536 p., in Russian.

Marty, B., Avice, G., Sano, Y., Altwegg, K., Balsiger, H., Hassig, M., Morbidelli, A., Mousis, O., and Rubin, M., Origins of volatile elements (H, C, N, nobble gases) on Earth and Mars in light of recent results from the ROSETTA cometary mission, *Earth Planet. Sci. Lett.,* 2016, vol. 441, pp. 91–103.

Morbidelli, A., Chambers, J., Lunine, J.I., Petit, J.M., Robert, F., Valsecchi, G.B., and Cyr, K.E., Source regions and timescales for the delivery of water to the Earth, *Meteorit. Planet. Sci.,* 2000, vol. 35, pp. 1309– 1320.

Morbidelli, A., Lunine, J.I., O'Brien, D.P., Raymond, S.N., and Walsh, K.J., Building terrestrial planets, *Ann. Rev. Earth Planet. Sci.,* 2012, vol. 40, no. 1, pp. 251–275.

Muralidharan, K., Deymier, P., Stimpf l, M., de Leeuw, N.H., and Drake, M.J., Origin of water in the inner solar system: A kinetic Monte Carlo study of water adsorption on forsterite, *Icarus,* 2008, vol. 198, no. 2, pp. 400–407.

Nesvorný, D., Roig, F., and Bottke, W.F., Modeling the historical f lux of planetary impactors, *Astron. J.,* 2017, vol. 153, no. 3, art. ID A103, 22 p.

O'Brien, D.P., Walsh, K.J., Morbidelli, A., Raymond, S.N., and Mandell, A.M., Water delivery and giant impacts in the 'Grand Tack' scenario, *Icarus,* 2014, vol. 239, pp. 74–84.

Pavlov, A.A., Pavlov, A.K., and Kasting, J.F., Irradiated interplanetary dust particles as a possible solution for the deuterium/hydrogen paradox of Earth's oceans, *J. Geophys. Res.,* 1999, vol. 104, no. E12, pp. 30725– 30728.

Petit, J.-M., Morbidelli, A., and Chambers, J., The primordial excitation and clearing of the asteroid belt, *Icarus,* 2001, vol. 153, no. 2, pp. 338–347.

Raymond, S.N., Quinn, T., and Lunine, J.I., Making other earths: Dynamical simulations of terrestrial planet formation and water delivery, *Icarus,* 2004, vol. 168, no. 1, pp. 1–17.

Raymond, S.N. and Izidoro, A., Origin of water in the inner Solar System: Planetesimals scattered inward during Jupiter and Saturn's rapid gas accretion, *Icarus,* 2017, vol. 297, pp. 134–148.

Rubie, D.C., Jacobson, S.A., Morbidelli, A., O'Brien, D.P., Young, E.D., de Vries, J., Nimmo, F., Palme, H., and Frost, D.J., Accretion and differentiation of the terrestrial planets with implications for the compositions of early-formed Solar System bodies and accretion of water, *Icarus,* 2015, vol. 248, pp. 89–108.

Sobolev, A.V., Asafov, E.V., Gurenko, A.A., Ardnt, N.T., Batanova, V.G., Portnyagin, M.V., Schonberg, D.G., and Krasheninnikov, S.P., Komatiites reveal a hydrous Archaean deep mantle reservoir, *Nature,* 2016, vol. 531, no. 7596, pp. 628–632.

Usui, T., Martian water stored underground, *Nature,* 2017, vol. 552, pp. 339–340.

Wade, J., Dyck, B., Palin, R.M., Moore, J.D.P., and Smye, A.J., The divergent fates of primitive hydrospheric water on Earth and Mars, *Nature,* 2017, vol. 552, pp. 391–394.

Yang, L., Ciesla, F.J., and Alexander, C.M.O.'D., The D/H ratio of water in the solar nebula during its formation and evolution, *Icarus,* 2013, vol. 226, pp. 256–267.

Zheng, X., Lin, D.N.C., and Kouwenhoven, M.B.N., Planetesimal clearing and size-dependent asteroid retention by secular resonance sweeping during the depletion of the solar nebula, *Astrophys. J.,* 2017, vol. 835, art. ID A207, 21 p.

*Translated by E. Petrova*